\documentclass{article}

 \usepackage[preprint]{neurips_2025}

\usepackage[utf8]{inputenc} % allow utf-8 input
\usepackage[T1]{fontenc}    % use 8-bit T1 fonts
\usepackage{hyperref}       % hyperlinks
\usepackage{url}            % simple URL typesetting
\usepackage{booktabs}       % professional-quality tables
\usepackage{amsfonts}       % blackboard math symbols
\usepackage{nicefrac}       % compact symbols for 1/2, etc.
\usepackage{microtype}      % microtypography
\usepackage{xcolor}         % colors
\usepackage{graphicx}
\usepackage{comment}
\usepackage{wrapfig}

\title{AI as intermediary in modern-day ritual:\\ An immersive, interactive production of \\ the roller disco musical \textit{Xanadu} at UCLA}
\author{%
  Mira Winick \\
  REMAP / Dept. of Theater\\
  UCLA\\
  Los Angeles, CA 90095 \\
  \texttt{mirawinick@ucla.edu} \\
  \And
  Naisha Agarwal \\
  Dept. of Computer Science\\
  UCLA\\
  Los Angeles, CA 90095 \\
  \texttt{naishaa@g.ucla.edu}
  \And
  Chiheb Boussema \\
  REMAP\\
  UCLA\\
  Los Angeles, CA 90095 \\
  \texttt{chiheb@g.ucla.edu}\\
  \And
  Ingrid Lee\\
  Dept. of CS / Theater\\
  UCLA\\
  Los Angeles, CA 90095 \\
  \texttt{ingridlee@g.ucla.edu} \\
  \And
    Camilo Vargas\\
 REMAP\\
  UCLA\\
  Los Angeles, CA 90095 \\
  \texttt{cjvargas@g.ucla.edu} \\
  \And
  Jeff Burke\\
  REMAP / Dept. of Theater\\
  UCLA\\
  Los Angeles, CA 90095 \\
  \texttt{jburke@remap.ucla.edu} \\
}

\begin{document}

\maketitle

\begin{abstract}
%Contemporary AI interfaces, including large language models, generative media pipelines, and perception modules, are often engineered for single user interaction. We investigate ritual as a scaffold for collective human–AI engagement through an immersive staging of the Broadway musical Xanadu at UCLA in Spring 2025. During a two-week run, about five hundred audience members contributed sketches and bodily movements that vision language models translated into responsive virtual scenery and choreographic prompts. Based on these observations and iterative design insights, we contribute (1) a ritual logic framework that recasts audience input as offerings and guides AI transformation of those offerings into performative media, (2) a meta-instrument design pattern that fuses sensing, generative AI and stagecraft so the entire theater ecology can be played as a single instrument, and (3) a reciprocity loop, exemplified by an Oracle sequence in the show where a Vision Language Model (VLM) generates poetic choreography that the audience performs, closing the human–AI feedback cycle. These findings show that AI can reinforce, rather than eclipse, group creativity and play, addressing a critical gap in prevailing single user AI design paradigms. 
Interfaces for contemporary large language, generative media, and perception AI models are often engineered for single user interaction. We investigate ritual as a design scaffold for developing collaborative, multi-user human–AI engagement. We consider the specific case of an immersive staging of the musical \textit{Xanadu} performed at UCLA in Spring 2025. During a two-week run, over five hundred audience members contributed sketches and jazzercise moves that vision language models translated to virtual scenery elements and from choreographic prompts. This paper discusses four facets of interaction-as-ritual within the show: audience input as offerings that AI transforms into components of the ritual; performers as ritual guides, demonstrating how to interact with technology and sorting audience members into cohorts; AI systems as instruments ``played'' by the humans, in which sensing, generative components, and stagecraft create systems that can be mastered over time; and reciprocity of interaction, in which the show's AI machinery guides human behavior as well as being guided by humans, completing a human–AI feedback loop that visibly reshapes the virtual world. Ritual served as a frame for integrating linear narrative, character identity, music and interaction. The production explored how AI systems can support group creativity and play, addressing a critical gap in prevailing single user AI design paradigms. 
\end{abstract}

\section{Introduction}

\begin{figure}[!h]
  \centering
  % \fbox{\rule[-.5cm]{0cm}{4cm} \rule[-.5cm]{4cm}{0cm}}
  \includegraphics[height=4cm]{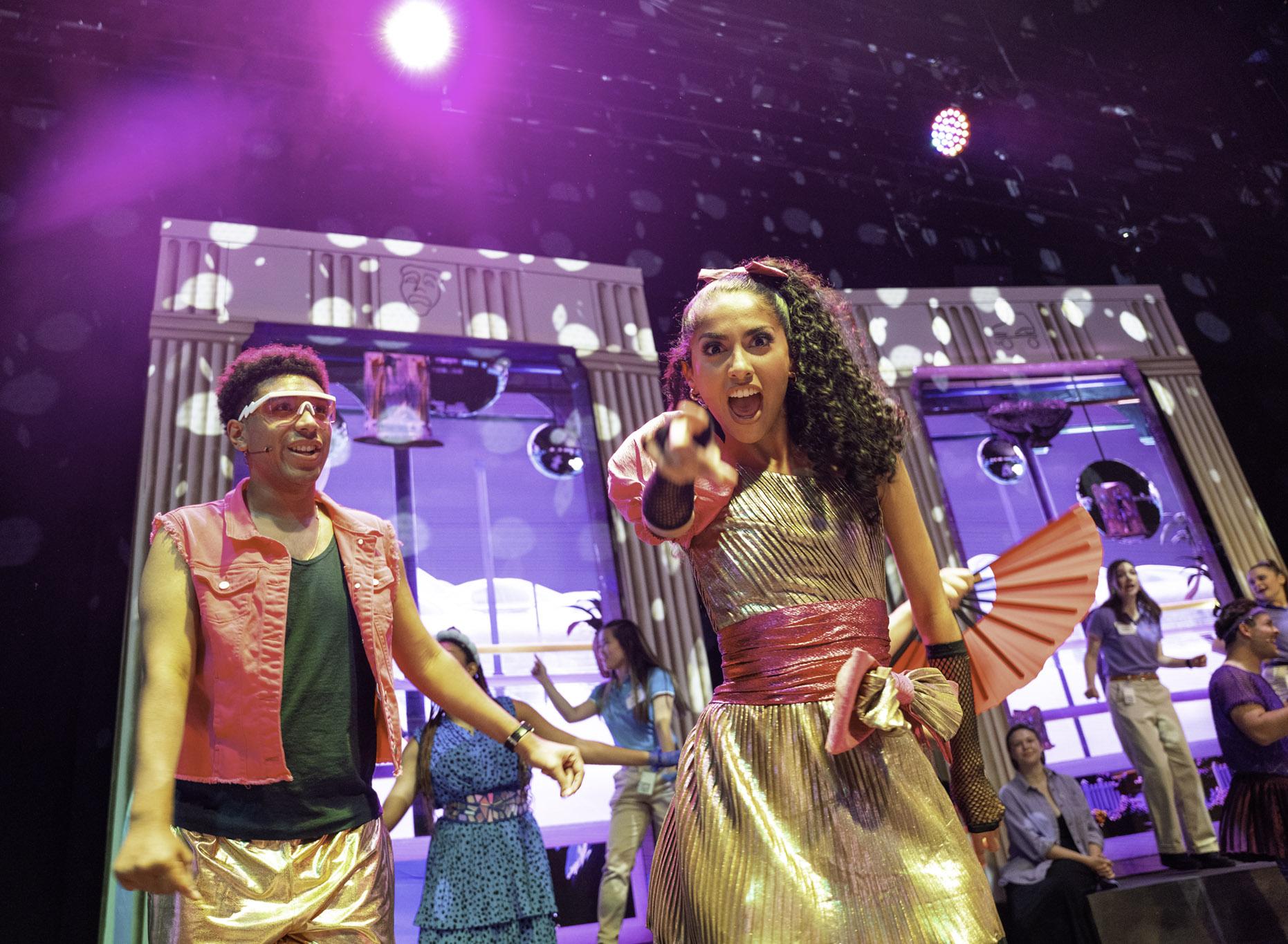}
  \includegraphics[height=4cm]{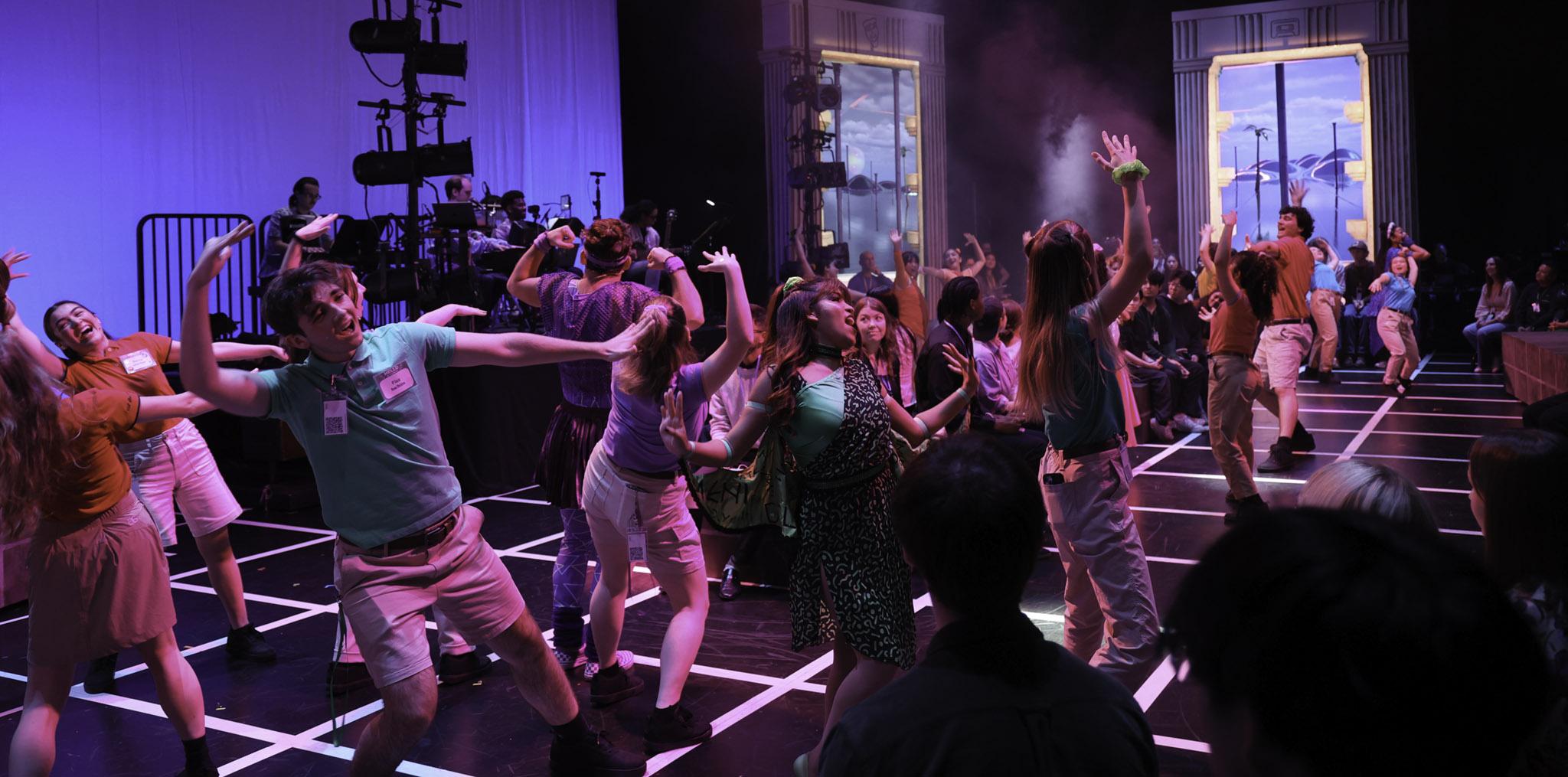}
  
  \caption{\textit{Xanadu} was an immersive musical performed at UCLA in May 2025.}
    \label{fig:intro}
\end{figure}

Mobile phones expanded the dominance of single-user interfaces for human-computer interaction from the ``personal computer'' era onwards. The emergent capabilities of LLMs and other foundation models to analyze and respond to a broad range of natural language and multimodal inputs (potentially from many sources at once) suggest new, AI-supported interface paradigms may be on the horizon and older, even ancient, concepts of human interaction may be more easily supported\footnote{There are notable explorations of group interfaces, from the tangible UIs of Ishii et al. \citep{ullmer2000emerging} to LLM-supported collaborative storytelling like Epic Saga Builder~\citep{epicsagabuilder}, but there are far more single-user interfaces.}. In this paper, we describe the use of contemporary AI techniques to enable group participation in an immersive staging of the Broadway musical \textit{Xanadu} at UCLA in Spring 2025\footnote{Presented through special arrangement with Music Theatre International (MTI).} (see Figure~\ref{fig:intro}). The production explored how AI-generated media from an audience could be integrated into designed extended reality (XR) scenery and sound. A key goal was to support collaborative creativity by audiences within an otherwise linear show that had a fixed narrative and musical spine. Conventional physical production elements (e.g., movable scenery, costumes, a live band, theatrical lighting) were integrated with their virtual XR counterparts (digitally realized scenery, characters and costumes, sound, lighting) rendered in real-time using Unreal Engine \citep{UnrealEngine5}. 
The show invited audience members to actively participate by making music, drawing sketches, and dancing together, with AI observing and then transforming contributions into aspects of the show’s evolving virtual world. 

% This paper focuses on how group audience participation through sketches and other types of input was encouraged and supported through generative AI components, enabling audience members to contribute to the show’s digital scenery and become part of its story of human creativity.\\
 One of the team's design strategies proved particularly effective for integrating audience input and its AI-driven transformations. We conceptualized the show as a \textbf{contemporary ritual}, in which technology functioned as a supporting mechanism enabling the uninitiated audience to form groups and join experienced performers in the ritual.
%Here, we focus on one design strategy that was particularly useful for integrating audience input and the resulting AI transformations into the experience: the conceptualization of the show as a \textbf{contemporary ritual}, in which technology acted as supporting machinery to enable uninitiated audiences to form groups and join the more experienced--performers--in that ritual. 
In this paper, we first introduce the show and its sources of inspiration. Then, we describe how ritual was used to design and understand group interactions, employing AI for transformation and synthesis of participation, and provide brief observations. 

\section{Production and research motivations}\label{sec:prod}

\begin{figure}[!h]
  \centering
  % \fbox{\rule[-.5cm]{0cm}{4cm} \rule[-.5cm]{4cm}{0cm}}

  \includegraphics[height=4cm]{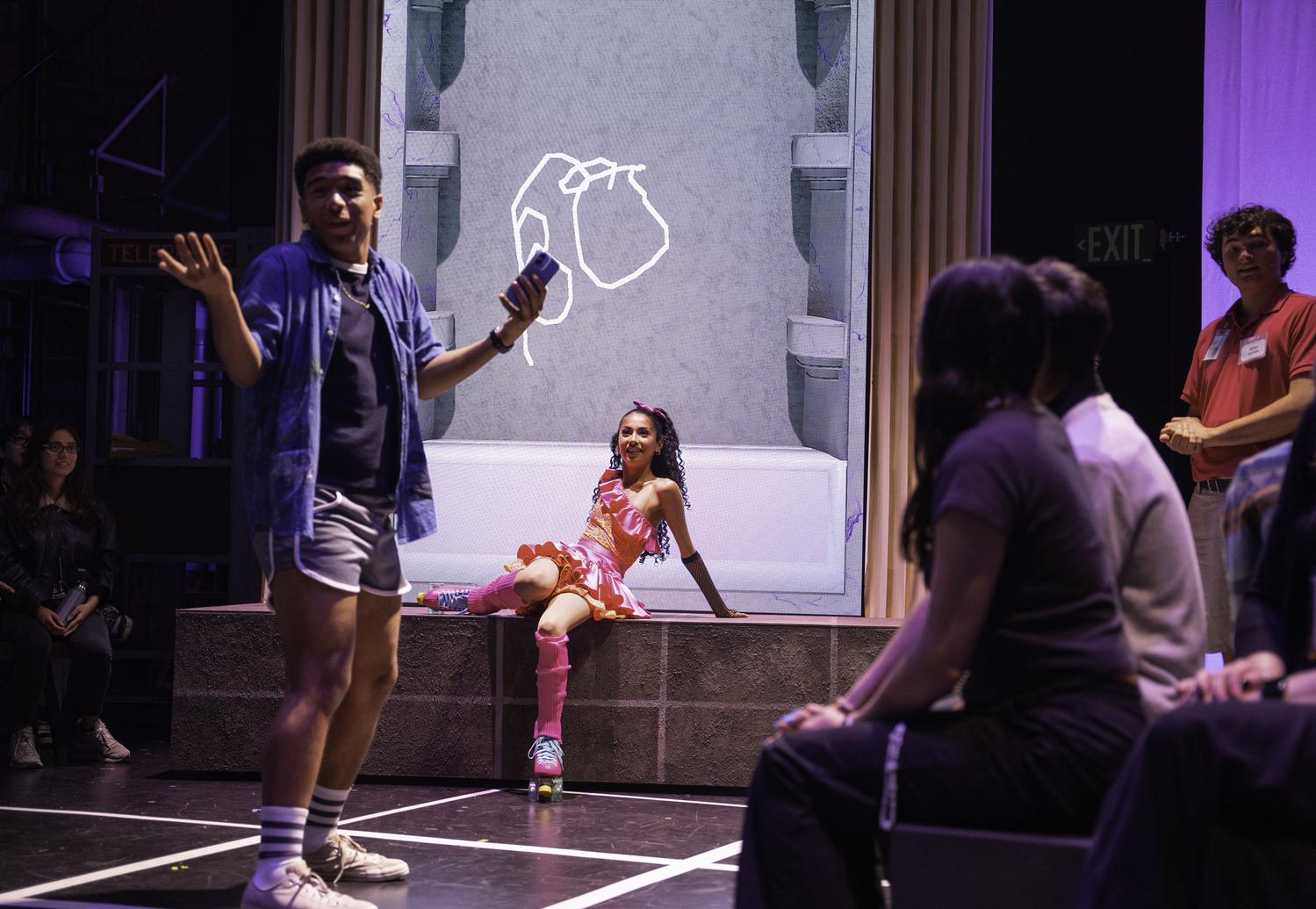}
      \includegraphics[height=4cm]{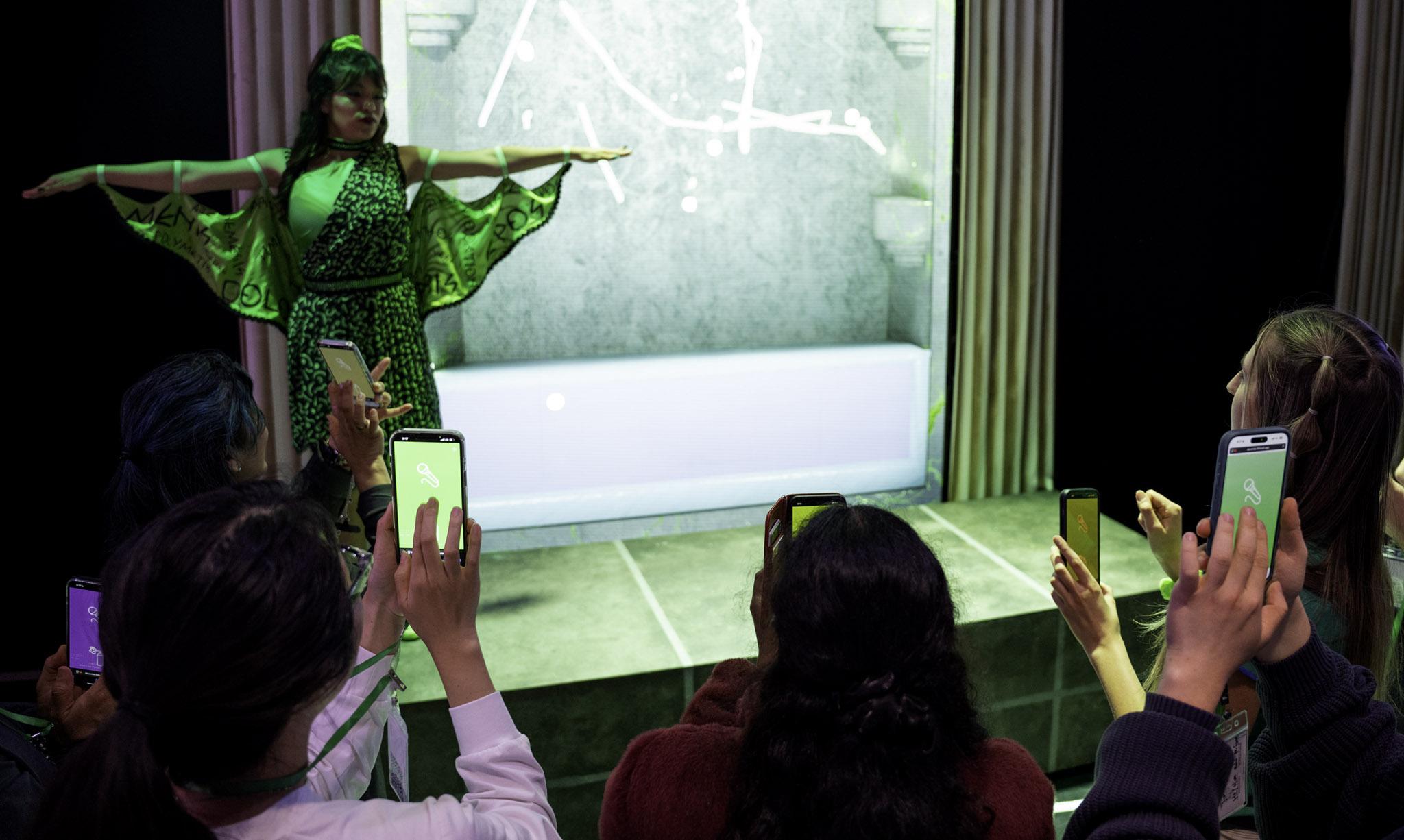}
  \caption{Audiences were shown by performers how to create drawings as offerings to the muses, which were transformed into images and objects  in the muses' ``shrines'' by generative AI.}
  \label{fig:shrine}
\end{figure}

 \textit{Xanadu}~\citep{xanadu} follows the Greek muse Clio, who comes to Venice, California in the early 1980’s as ``Kira'', to inspire the mortal artist Sonny to pursue his creative dreams and open a roller disco. Over about a year, the UCLA Department of Theater and UCLA REMAP developed  an immersive, participatory staging of the complete Broadway musical, performed for public audiences in May 2025. In this unique production, the audience was seated onstage, surrounded by seven thirteen-foot high, movable LED displays---\textit{shrines}. Each shrine  corresponded to one of the show's seven muses, including Kira / Clio. Audiences played an important role in Sonny and Kira’s journey by interacting in person and contributing their own creativity to the virtual world and the resulting roller disco brought to life around them. 

We\footnote{The authors include the faculty creator / producer (Burke), one of the faculty directors (Winick), the assistant choreographer (Lee), and AI / systems developers (Agarwal, Boussema, Lee, Vargas, Burke).  
%Full credits are at \url{xanadu.remap.ucla.edu}.
% also in acknowledgements
}  embarked on the production to investigate how AI and XR could create novel audience experiences. An early design decision was to frame the show as a modern-day ritual, paralleling aspects of Greek ritual, which is both well known and directly connected to the show's narrative. The team followed this frame to work coherently across the real-world and fictional interpretations of any given moment. Other theatrical traditions, particularly those involving participation and improvisation, such as Commedia dell'Arte \citep{oreglia_commedia}, provided additional tools and perspectives.   
Greek rituals often feature a large gathering of lay participants guided by ritual ``experts'' to create and perform offerings for a non-human entity \citep{parker2011greek}. \textit{Xanadu} reproduced this structure: the audience became the ritual crowd, led by an ensemble of ``acolytes'' to participate with the lead performers, including in the creation of digital offerings presented to the muses. Participants began in a liminal space between the real world and the show's world, where acolytes "onboarded" them into muse groups. The acolytes acted as guides for the audience throughout the show, but also fluidly crossed the "fourth wall" to serve as a traditional ensemble, participating in the show's songs and choreography. 
% They prompted audience members to contribute sounds, sketches, gestures, and movements that are then interpreted by AI systems and rendered back into the virtual world.  

We elaborate on specific design strategies below. Engaging with collective participation explores a gap in current conceptualizations of user interaction with contemporary AI systems. It shifts the focus from predominantly individual interfaces to AI to emerging models of group-based interaction \citep{lee2025beyond}. 
The default focus on personal interfaces can be traced to early user experience (UX) research and cognitive psychology, which often emphasized individual-level performance metrics—an emphasis that continues to shape product design, human–computer interaction, and, more recently, human-centered AI \citep{lee2025beyond}.
%and the predominantly personal interaction focus of UI design in general. 
% While much human interaction occurs in groups, most AI systems continue to assume a single-user paradigm. 
Frameworks for collective engagement with AI remain comparatively underdeveloped. 
%especially in creative production contexts like this one. 
A growing body of work recognizes the need for more research on interactions with AI in groups 
%\citep{shiiku2025dynamics, cui2024ai, naiseh2024xai, nigro2025social, muller2024group} 
\citep{cui2024ai, shiiku2025dynamics, borghoff2025human, lee2025beyond}
as these present unique challenges, including issues of perceived ownership \citep{zhang2025exploring},
effective interaction modalities \citep{raees2024explainable}, user engagement and agency \citep{zhang2025exploring, raees2024explainable}, trust, group and power dynamics, the navigation of varying levels of experience, knowledge,  preferences \citep{naiseh2024xai}, as well as unpredictability-related trust issues and frustrations %that characterize generative AI and contrast with more traditional 
with non-determinism \citep{xu2023transitioning}. 
%In our work, we overcame several of these challenges by helping to create a shared mental model between the humans and AI within a ritual framework as we explain in the next sections.

% Among this growing body of investigative work, we distinguish those focused on intellectual tasks and collective intelligence \citep{cui2024ai, naiseh2024xai} such as brainstorming \citep{muller2024group} and those focused on creative applications such as storytelling \citep{shiiku2025dynamics}. Our work contributes to the latter category as we present our conceptual architecture framing group-level interactions with AI in a manner that emphasises human collaboration and creativity and allows for the AI output to reflect the communal creativity that is its input. 

% In our work, we addressed several of these challenges through the ritual framework adopted to contextualize the group-AI interaction within the theatrical production as we explain below.
Engaging these challenges through artistic production, our work adopted a ritual framework that situates group-AI interaction within a theatrical performance that needed to operate in both fiction and reality at once, offering a means to explore collective participation, trust, agency, and play.

% \section{Facets of performance ritual framework}\label{sec:ritual}
\section{Designing for audience participation in a performance ritual framework}\label{sec:ritual}

Embedding audience participation within a theatrical production that must follow strict script and timing constraints is not trivial. Some of the challenges that must be addressed involve (i) how to inscribe the audience's participation within the narrative flow, (ii) how to demonstrate the interaction modalities on-the-fly and in-story, (iii) how to get the audience engaged and willing to contribute in front of everyone else, and (iv) what to do with the aleatory and unpredictable aspects of generative AI. These accompany technical implementation challenges related to performance requirements and budget constraints, as well as a broad need to ensure generated media are aligned with the show's aesthetics. For the interested reader, we touch on the implementation challenges in Appendix~\ref{appendix}.

% \subsection{Interaction modalities}\label{subsec
% :modalities}

\subsection{Input as offerings transformed by AI}\label{subsec:offerings}
A quintessential element of Greek ritual is the offering of gifts by mortals to the gods
%, not as a direct exchange, but
as an invitation for divine favor. This practice is exemplified by the festival of Chalkeia, where Athenian women collectively wove sacred robes as an offering to Athena, blending artistic labor with devotional intent \citep{parker2011greek}. In \textit{Xanadu}, all audience contributions, of drawing, music, and dance, were conceptualized as offerings, with AI as an intermediary translating these gifts into more polished representations in the virtual world inhabited by the show's gods. These offerings of phone-based drawing, sound-making, and skeletal tracking of movement (see Appendix \ref{appendix} for more details) underscore the opportunity to expand interaction modalities in XR beyond speech, text, and gesture~\citep{raees2024explainable}. Phone positional tracking enabled participants to draw on the shrine-canvases with their phones as wands, with generative AI transforming their sketches into polished images and 3D assets (see Figure~\ref{fig:shrine}). Skeletal tracking of dance and gesture, using computer vision, enabled group-level movement-based communion (see section ~\ref{subsec:reciprocity}).
 
The ritual aspect framed audience participation consistently with the plot and characters while establishing legible norms and conventions for design, rehearsal, and eventually audience contributions. \textit{Improbotics} \citep{mathewson2018improbotics} used a similar approach, in which theater performers integrated lines from generative AI while performing. Lines were viewed as ``offers" for an actor and, with the cast's collective engagement, were integrated into scenes being performed. \textit{Xanadu} worked in the reverse: performers prompted the audience, audience members contributed responses, and AI interpreted these to generate changes in the virtual world. Here, the human acolytes guided improvised exchanges with the generative AI, mirroring historical rituals where expert practitioners instructed participants in producing spontaneous offerings devoted to particular gods or ritual themes \citep{parker2011greek}. Merging ancient models of guided offering with AI-driven improvisation, the show used ritual logic to coherently structure collective human–machine interaction.

\begin{figure}[!h]
  \centering
  % \fbox{\rule[-.5cm]{0cm}{4cm} \rule[-.5cm]{4cm}{0cm}}
  \includegraphics[height=4cm]{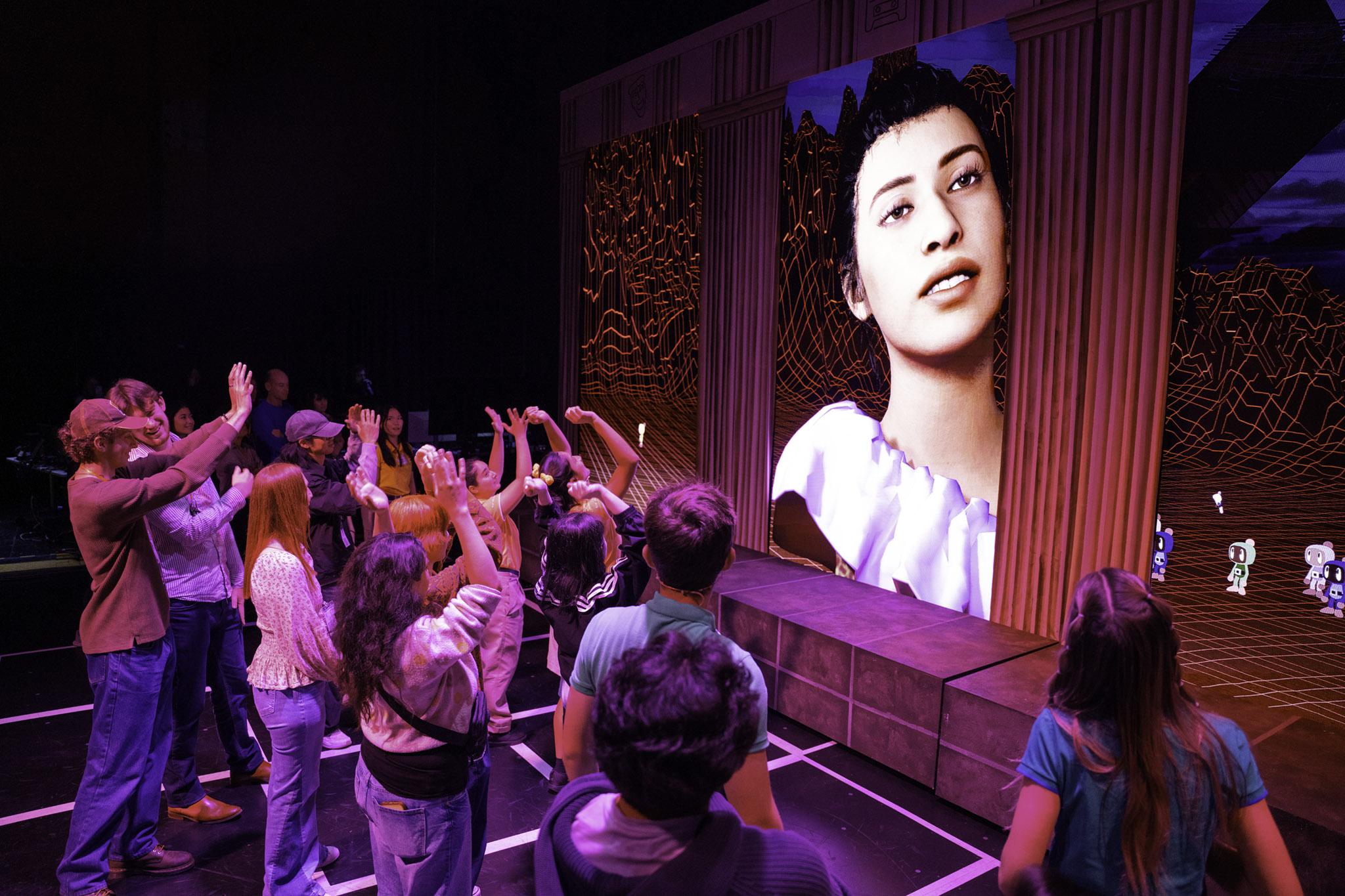}
  \includegraphics[height=4cm]{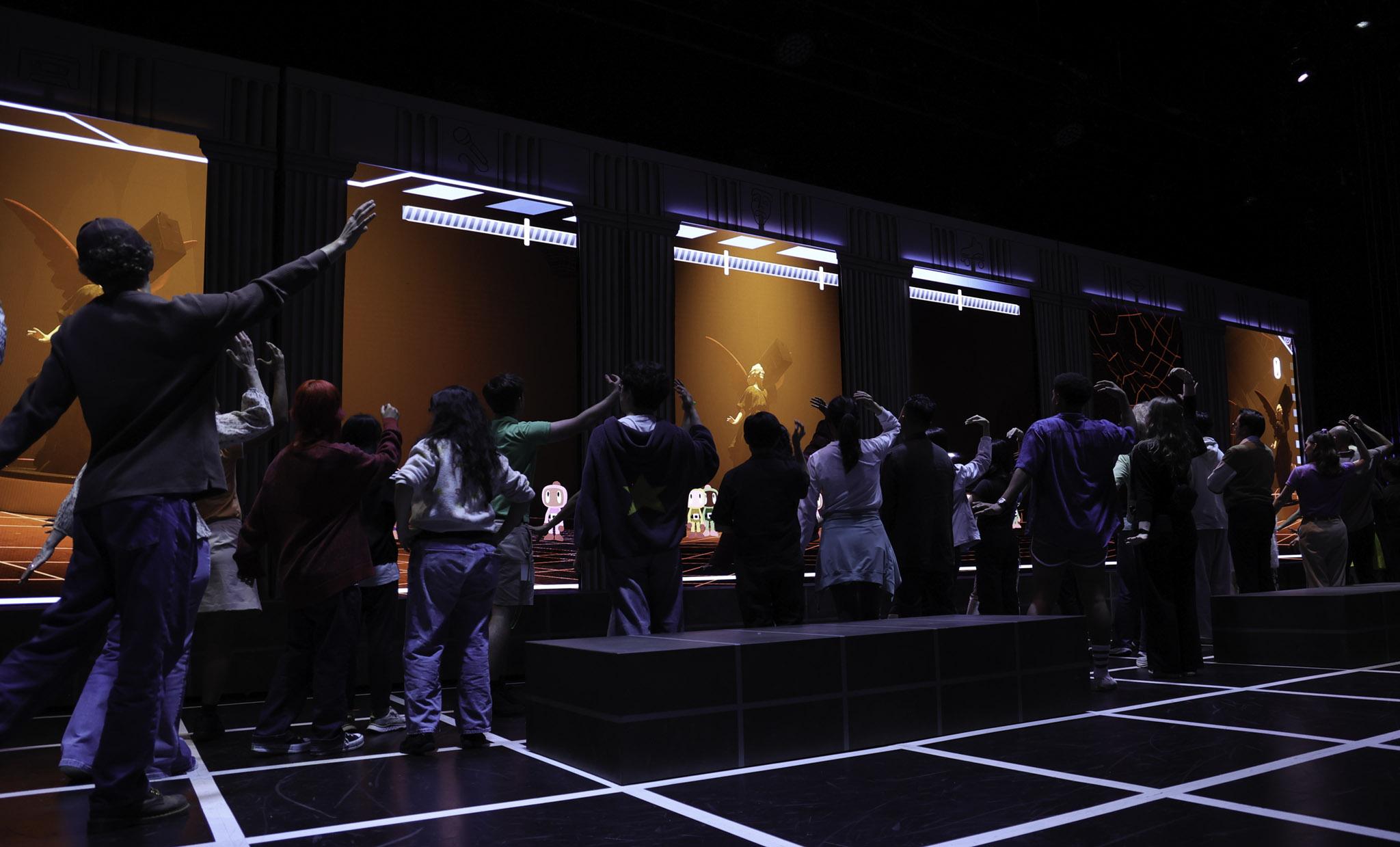}
  \caption{In Olympus, the audience's movement ultimately brings Kira and Sonny back together.}
    \label{fig:olympus}
\end{figure}

\subsection{Invitation to interact via performer demonstration}
\label{subsec:invitation}
During the song “Magic”, Sonny first sketches Kira using a phone borrowed from an acolyte, and then invites the audience to sketch ideas for his roller disco. Following the ritual framework, acolytes drew first, acting as expert participants to demonstrate the process. This allowed them to show examples of valid inputs and model collaborative drawing, signaling to the audience that even rough scribbles were welcome. The focus on collective contribution over perfection lowered audience performance anxiety and instead encouraged collaboration. Since large language and diffusion models still struggle with contextual nuance and visual consistency \citep{avrahami2024chosen}, we chose to embrace these limitations instead of trying to hide them.
%, foregrounding interpretation over technical precision.

Through three successive cycles of drawing and generation, each modeled by the acolytes, the audience developed their own expertise with the drawing instrument.  Generative AI pipelines, composed of vision-language and diffusion models (see \ref{appendix:sketching}), converted each sketch into an image or a 3D mesh satisfying the show's vaporwave aesthetic, muse-specific color palettes, and overall artistic direction. Similar to collaborative art installations like Dream Painter \citep{guljajeva2022dream} and FRIDA \citep{schaldenbrand2022frida}, collective input guided the AI system, which produced elements for a now audience-expanded virtual world. For instance, when a spectator sketched a roller skate, the AI (generally) generated a 3D asset of a skate rendered in the show's vaporwave style. In cases where a generation misfired, a human‑in‑the‑loop moderator could intercept the output. The acolytes explained remaining oddities as ``mysterious insight" into the audience's intent. This dramaturgical structure re-framed algorithmic opacity and unpredictability as  %$capricious wisdom of a higher power 
part of the storyworld's metaphysics. LuminAI \citep{trajkova2024exploring} used similar tactics, treating an AI dance partner’s unexpected moves not as mistakes, but openings for improvisational responses and dialogue. 

Furthermore, to strengthen collective identity, each audience group (about eight people plus two acolytes) was dedicated to one Muse. They had their own shrine with a unique iconography and color scheme. The AI output for each shrine reflected the perspectives and journeys of the associated group, shaped by the influence of their Muse. To achieve this, we took a two-fold approach. First, acolytes provided drawing cues aligned with their particular patron. For example, the Muse of Music’s group needed to collaboratively sketch musical instruments on their shared canvas. Second, the AI pipelines were internally directed towards each Muse's aesthetics using designer-selected color palettes and reference imagery. Acting as a unifying force, the AI systems wove individual contributions into cohesive outputs embodying small group identity within the larger whole.

\subsection{Stage machinery (including AI) as ritual instrument}\label{subsec:machinery}
Inspired by the Eleusinian mysteries, where initiates’ song and dance summoned the gods’ realm \citep{parker2011greek}, \textit{Xanadu}'s virtual world was conceived as a parallel sacred space accessed by the ritual participants through the shrine set pieces. The relationship between the physical and digital worlds was similar to Fiebrink’s concept of the “meta-instrument”, in which controllers are taught to convert sensed inputs into a bounded space of outputs, creating a new instrument to be played \citep{fiebrink2016machine}. In that work, controllers are specific ML processes. \textit{Xanadu} employed a broader network of relationships transforming the input of human physical actions into XR outputs via various AI-supported systems. For example, positional tracking of the shrines enabled performers to move them to change the LED screens' perspective into the virtual world, without technician intervention.
% The muse-performers emerged and returned through these shrines, moving between the physical stage and the virtual world they inhabited.
%One of the key instruments enabling this ritual exchange was the audience’s phone, which, through a custom WebXR tool, mapped physical input to virtual output, transforming gesture into digital offerings visible within the virtual world.\citep{parker2011greek}.
A WebAR app on audience phones, acting as a ritual instrument, transformed  gestures (made with the whole phone) into offerings. In groups, the audience aimed their phones at the shrines
%, pressed a finger to the screen, 
and, moving them as paintbrushes, drew  %as its SLAM process captured the motions 
  colored poly‑lines in the virtual world. (See \ref{appendix:overview}.)  One of three custom AI pipelines, used at different moments in the show, processed the inputs. (\ref{appendix:sketching}) 
  %, selected by the actor’s prompt. 
  A vision–language model, coupled with diffusion models, merged the audience sketches with designer reference images and actor portraits, producing 2D images or 3D meshes matching the show’s aesthetics. 

%selected by the actor’s prompt. A vision–language model, coupled with diffusion models, merged the audience sketches with designer reference boards and actor portraits, producing 2D images or 3D meshes aligned with the show’s aesthetic palette. These interactions, interpreted as ritual acts, embodied the meta-instrument concept in which human input, sensing technologies, and AI transformation operated as a single expressive system.

To develop the relationship between audience input and digital output, the team conducted weekly play-tests with invited participants during the show's development and rehearsals. Through this iterative process, designers adjusted gesture sensitivity, stroke width, emissivity, and color parameters, and refined model weighting to align the outputs with the production’s aesthetics. By the performance, the system had a single gestural grammar for the audience, customized aesthetics for each Muse group, and different AI pipelines for specific moments. Actors prompted Muse image generation in one scene and 3D object creation in another, with audiences performing the same gestures for their own results. This gave the audience one interface to learn to participate in several aspects of the ritual. %This consistent yet flexible mapping preserved audience ease and gave the creative team control over thematic coherence and visual variety.

%Weekly play‑tests let designers tune gesture sensitivity, stroke width, emissivity, and color rules, while iterative weighting of model inputs gradually aligned the outputs with the show’s aesthetic. Maintaining a single gestural grammar but swapping pipelines on the back-end allowed actors to request muse image generation in one scene and 3D objects in another, giving spectators varied results without requiring them to learn a new interface. This flexible yet consistent mapping preserved audience ease while granting the creative team fine control over thematic coherence and visual variety. The ritual framework enabled the meta-instrument relationship between physical interaction and the virtual world, shaping how audience groups engaged with the system and legitimizing exploratory collaboration.

In the final act, the Muses and the mortals--Sonny, the acolytes, and the audience--ascended to a virtual Olympus (see Figure~\ref{fig:olympus}). Computer vision (CV) tracked the mortals' position and pose using off-the-shelf facial and body detection with ML. This data drove Bomberman-inspired avatars \citep{superbombermanR}, whose puny movements in the virtual world contrasted with the large area covered by the audience on stage. %that moved widely across the physical stage yet advanced only marginally in the virtual world. 
Actors playing gods stood on a narrow runway, visibly piloting larger-than-life virtual versions of themselves. Both muse and mortal bodies played the virtual avatars as instruments, with CV-based stage machinery translating their movement into an active manifestation of hierarchy and power within the virtual world. The integration of ritual logic with the meta-instrument framework helped the team to design a cohesive relationship between physical gestures, sensing, AI processes, and corresponding digital outputs, and a unified grammar for integrated audience participation.

\subsection{Reciprocity: AI-directed audience choreography}\label{subsec:reciprocity}

% created a reflexive loop between digital and physical domains, allowing subtle tweaks, such as adding spatial set shifts to the breath instrument, to reverberate across the entire staging. 

\begin{wrapfigure}{r}{0.4\textwidth}%[!h]
  \centering
  % \fbox{\rule[-.5cm]{0cm}{4cm} \rule[-.5cm]{4cm}{0cm}}
  \includegraphics[width=0.38\textwidth]{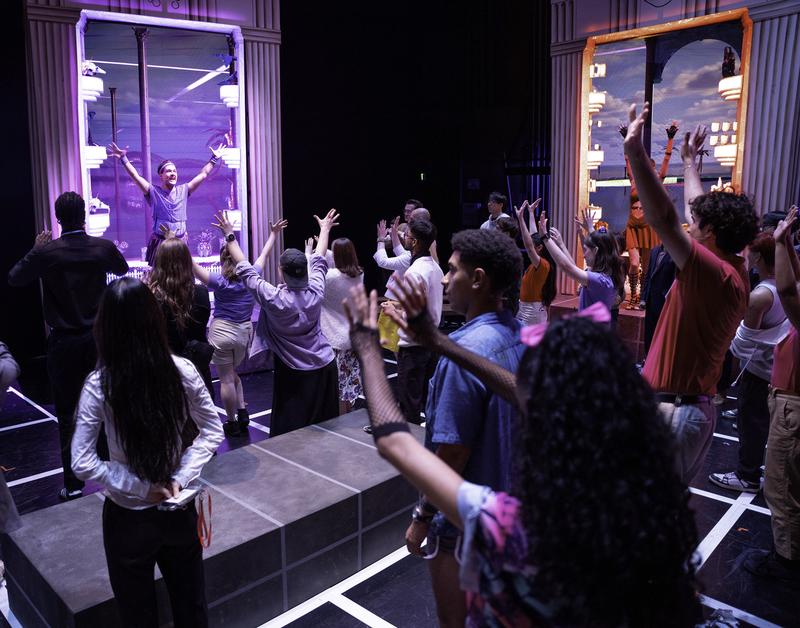}
  \caption{The audience's dance moves help build the \textit{Xanadu} auditorium.}
    \label{fig:choreo}
\end{wrapfigure}

Prior sections discuss a unidirectional structure: human action $\rightarrow$ AI/ML process $\rightarrow$ media output. With the final offering of dance, as shown in Figure~\ref{fig:choreo}, we reversed this flow. An AI component guided audience movement, creating a deliberate reciprocity with early scenes. 
This echoes role reversals in Greek rituals, which temporarily invert hierarchies~\citep{parker2011greek}. In the norms of Greek ritual, the gods do not speak directly to mortals. Messages pass through intermediaries who reveal divine instructions~\citep{parker2011greek}. An actor called the Oracle consulted an iPad with access to a vision–language model (VLM). The VLM analyzed that performance's sketch-driven, AI-generated media offerings and output a short poem. It used that poetry to describe three jazzercise-style moves selected from a curated set. The Oracle read the poem aloud and, with the Muse of Dance, taught the moves to the audience. With the acolytes enthusiastically demonstrating, the audience was invited to perform the choreography. A depth camera rig tracked skeletal poses and compared collective motion to a baseline. If the crowd’s energy, timing, and alignment exceeded a threshold, the virtual environment could respond dynamically with lighting shifts and scene updates. In this way, we explored how immediate engagement of the whole audience could impactfully contribute to the show.  

%To evaluate audience participation in this performative interaction layer, a computer vision pipeline using pose estimation tracked and analysed individual and group movement. This skeletal tracking system calculated the position and velocity of 2D joint coordinates, enabling a multimodal tracking framework that allowed for real-time feedback and adaptivity, while also enabling dramaturgical decisions based on audience energy and synchronicity.

\begin{comment}
\begin{figure}[!h]
  \centering
  % \fbox{\rule[-.5cm]{0cm}{4cm} \rule[-.5cm]{4cm}{0cm}}
  \includegraphics[height=4cm]{figures/dance1.jpg}
  \caption{The audience's dance moves help build the Xanadu auditorium.}
\end{figure}
\end{comment}

%The jazzercise choreography presented by the Oracle thus functions as part of the overall ritual that emphasizes the entire audience's shared identity, merging all the individual subgroups into one and transitioning audience members into co‑performers. As audience members learn and perform the selected dance moves, this choreography becomes symbolic and communally significant. Empirical studies have shown that moving in rhythm with others, particularly in contexts involving moderate physical effort, can increase interpersonal trust, emotional closeness, and feelings of affiliation \citep{tarr2015synchrony}. Enacted collectively in real time, they become rituals of belonging, reinforcing group unity through synchronized motion and creating space for shared embodied memory. 

The choreography, presented as ritual, merged audience subgroups into a whole via the VLM analysis, turning spectators into co-performers. Research shows that synchronized motion, especially with moderate effort, improves trust, emotional intimacy, and group affiliation \citep{tarr2015synchrony}. As participants danced together, movement became symbolic and communally significant.  Shared movements became rituals of belonging, reinforcing audience group unity through embodied memory.
%The enactment effect, which demonstrates that physically performing meaningful gestures significantly enhances memory encoding and recall compared to passive observation or verbal rehearsal \citep{roberts2022enactment}, highlights how the repetition of these choreographed sequences in \textit{Xanadu} not only helped the audience retain the elements of the story, but also anchored them in a shared performative context. Informal feedback from audience members supported this connection: when asked to reflect on memorable aspects of the experience, many cited the participatory dance as a highlight, indicating both enjoyment and a sense of engagement working toward a common goal. The Oracle sequence thus exemplifies how AI, when embedded in ritual dramaturgy, can reinforce group belonging and shared creative experience, offering a compelling counterpoint to dominant single-user AI interaction paradigms.
The enactment effect, where physically performing gestures improves memory retention over passive observation \citep{roberts2022enactment}, further suggests how repeated choreography in \textit{Xanadu} anchored story elements in shared experience. Informal feedback confirmed this: many audience members cited the participatory dance as a highlight, reflecting enjoyment and a sense of shared purpose. The Oracle scene thus illustrates how AI systems, when embedded in ritual dramaturgy, can foster group belonging and creativity, challenging the dominant single-user AI paradigm.

\section{Limitations and societal impact}\label{subsec:limitations}
%Because of the nature of Xanadu and it being a live performance, all audience data presented here is  anecdotal, based on the authors observations throughout the process. We believe Xanadu provides valuable insights into integrating technology into live performance, and provides a novel framing for how to view group-AI interactions.

This work demonstrates how historical understandings of ritual could guide group interaction with AI, but several limitations remain. Notably, the approach was tailored to a specific musical, creative approach, and physical space. Reproducibility was not a focus. 
% While we describe our system implementation and software architecture which can inspire similar work, the approach may not be entirely reproducible in other settings. 
Assessment of audience experience was informal and observational, limiting evidence for any claims of generality. While considerable technical effort was made to reduce generative AI latency to meet the show's interactivity goals, 
%(seven generations in less than a minute), 
 group-AI improvisation, especially involving diffusion models, remains constrained by latency, quality, and budget.
%Additionally, we may discuss potential broader societal impacts. 
Finally, while framing AI not as a god but at least as divine machinery served the show's goals, it risks reinforcing problematic perceptions of AI as authoritative, inscrutable, or all-knowing. 
%THe theatrical metaphor, though playful, may obscure the underlying 
Human authorship and curation, while present, are explained within the fiction rather than directly, though we did offer various behind-the-scenes conversations to the public. 
%, and limitations of generative systems—particularly for audiences unfamiliar with AI’s constraints. 
Finally, although we aimed to  
support collective creativity and co-authorship, generative models introduce worrying homogenizing tendencies \citep{doshi2024generative, anderson2024homogenization}. Their outputs often reflect dominant cultural aesthetics, tropes, and norms embedded in training data, which reduce diversity and nuance. The production tried to mitigate this through curated prompts, human moderation, and stylistic oversight, %future uses of similar systems in participatory settings should be 
but we remain mindful and concerned about the erosion of creative diversity. % whose voices, images, and narratives are being reproduced—and whose are being erased.

\section{Conclusion: generalizations and future impact}
Theater appears to mimic aspects of the ``real world'', so it can be tempting (and, at times, helpful) to think of it as a bounded testbed for new interfaces and system designs, tested within the microcosm of a given performance. Theater is also a unique human activity, with many different forms and historical antecedents. The types of ritual cited here are just a few of many from across the world. Each performance establishes and operates within its own rules for relationships among audiences and performers, fiction and reality, and humans, design, and technology. Strategies employed in performance find their way into film and television, gaming, and social media. They are also part of our everyday lives. Thus, this exploration of ritual framing to design and implement group interactions with AI may have utility for a range of collective group-AI interactions.  In future work, we plan to examine more deeply the role of rehearsal as a site for training, experimentation, and refinement of human–AI interaction. We are also interested in how AI can support small group interactions within much larger wholes, by enabling tailored interactions with global-scale narrative worlds, scaled through decentralization of AI onto edge devices. Generally, we seek AI-supported methods that engage live audiences in new forms of collaborative fandom and collective meaning-making.

\subsection*{Acknowledgements}
We thank the production's many contributors. Full show credits are at \href{https://xanadu.remap.ucla.edu/}{xanadu.remap.ucla.edu}. Photos in this article are by M. Yepez, M. Beymer, and K. Liu. In-kind equipment and services were provided by Boxx, Mo-Sys, and 4Wall Entertainment. Funding was provided by the UCLA Department of Theater, UCLA DataX, the Amazon MGM Studios' Innovative Storytellers Initiative, and Qualcomm.

% This paper demonstrates how ritual dramaturgy, meta‑instrument design, and bidirectional language workflows can work together to broaden non‑expert participation in AI‑driven theater while preserving narrative coherence. 

% [Recap facets here] 

% Mapping ritual elements onto Xanadu transformed music‑making, sketching, and dancing into recognisable “offerings” and reframed algorithmic unpredictability as divine caprice. Viewing the entire stage ecology as a single meta‑instrument supplied a shared design grammar (sensor $\rightarrow$ mapping $\rightarrow$ output) and let our team iterate quickly across disciplines, turning physical gestures into coordinated shifts of sound, light, and virtual scenery. The Oracle sequence further showed that AI can both read and direct human action.\\
% Together these strategies suggest a reusable approach for live performance: use ritual logic to scaffold group participation and treat interactive subsystems as facets of an instrument. \\
% We are currently exploring how to expand this ritual‑instrument framework from handcrafted rules to scalable, data‑driven practice and paving the way for more responsive, personalised, and adaptive theatrical worlds.  (maybe allude to fact-to-fiction mapping briefly)

%\newpage
\bibliographystyle{apa}
\bibliography{references}

\newpage
\appendix

\section{Appendix/Supplemental Material}\label{appendix}

\subsection{Brief overview of the AI-enabled show proceedings}\label{appendix:overview}
We briefly describe technical details of how a typical performance of \textit{Xanadu} proceeded, focusing on the key AI-enabled audience participation moments, taken in breaks between scripted material. 

As audience members arrived, they were divided into seven groups, each group corresponding to one of the seven muses in the story. Two "acolytes" (performers) guided each group through the various interactions in the show. Upon arrival, audience members were prompted to scan a QR code to access our web app on their phones, before taking their seats in the center of the stage. 

During the opening sequence, participants were invited to contribute to their muse's shrine by creating sketches at three key moments, each with a different creative task: (a) designing a background image for their muse, (b) drawing their muse in a pose of their choice, and (c) creating objects to populate the shrine. Phones acted as gestural interfaces; a WebAR app, built on the 8th Wall platform, translated 6DOF tracking and pose estimation into sketch strokes. 

These sketches became the starting point for our generative AI pipelines. Background sketches were transformed into high-quality images with the muse composited as a "frieze" at the bottom of the shrine. Pose sketches generated fully clothed, realistically proportioned muse characters that preserved facial identity and matched the designer's aesthetic. Object drawings were turned into 3D assets, that were then placed within the virtual shrine environment. Further details on these pipelines can be found in \ref{appendix:sketching}.

Beyond sketching, the show incorporated additional AI-driven segments that extended audience participation into new modalities. In the Oracle sequence, a vision-language model (VLM) analyzed a show's sketch-driven generated media, composed a short poem, and selected three dance moves from a curated set of low-impact, jazzercise choreography. In the final act, mortals and gods ascended to a virtual Olympus with the seven shrines being assembled into a continuous LED wall, and gods re-embodied as life-sized MetaHuman avatars driven by live motion capture (depth camera + iPhone LiveLink). 

% % Led by the actors, the audience performed these moves in unison. A computer vision model tracked skeletal poses and compared collective motion to the actors’ baseline, triggering dynamic environmental changes such as lighting shifts and scene updates.

% The final act brought technologies together as mortals and gods ascended to a virtual Olympus. The seven shrines were assembled into a continuous LED wall, and the gods (some of whom had just appeared in mortal form) were re-embodied as life-sized MetaHuman avatars driven by live motion capture (depth camera+ iPhone LiveLink). Mortals, including actors and audience members, were represented as squat bomberman-style figures inferred solely from depth silhouettes. The motion capture was staged openly; actors playing the gods stood on narrow runways to pilot their avatars, while mortals moved freely across the entire stage to make incremental progress in the virtual realm. This culminated in a lightning-bolt mini-game, where spectators stepped forward and back to shield the protagonist from Zeus's projectiles.

Code for these various modules can be found at our Github page \url{https://github.com/remap} and upon request.

\subsection{Sketching: generative AI pipelines}\label{appendix:sketching}

To support the interactive sketching tasks, we combined  contemporary AI methods, in particular multimodal large language models (MLLMs) and diffusion models, within a hybrid model hosting architecture. Some models were deployed directly on custom Amazon Web Services (AWS) SageMaker endpoints, giving us full control over model architecture, inference parameters, and generation logic. Others were accessed via AWS Bedrock, which provided faster inference for large models, albeit with limited configurability. 

Across the three generative tasks, the pipelines followed a common structure. MLLMs interpreted audience inputs and generated prompts while local diffusion models, augmented with state-of-the-art control techniques, generated task-specific media outputs. Occasionally, Bedrock-hosted diffusion models were used for quality enhancements. Detailed description of each task follows. 

\textbf{Generation task \#1: a scene for each muse}

\begin{figure}[!h]
  \centering
  % \fbox{\rule[-.5cm]{0cm}{4cm} \rule[-.5cm]{4cm}{0cm}}
  \includegraphics[scale =0.37]{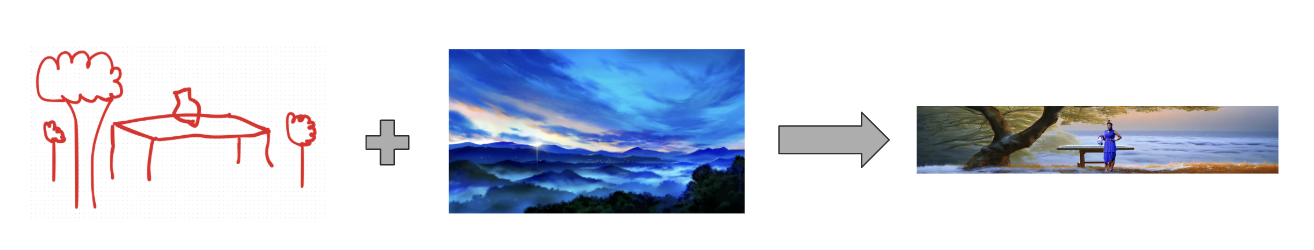}
  \caption{Generation Task 1, a scene for each muse}
\end{figure}

\textit{User Experience: } In this generation task, audience members (in pairs) drew a sketch of a background they wanted their muse to be seen in. That sketch was then turned into a generated background with the muse composited on top, placing each muse in a new environment. Designers, in addition to crafted prompts that fit their aesthetic design, provided a per-muse scene reference image to steer both the color palette and the type of background each muse would appear in. Performers (both muses and acolytes) would adapt to the scene that was generated for them, letting audience members almost transport their character into a new world.

\textit{Implementation: }The audience background sketch is passed into the DeepSeek Vision Language Model (VLM) ~\citep{lu2024deepseekvlrealworldvisionlanguageunderstanding} to generate a detailed text description of the image. We pass this text description, a designer style reference image, and the audience background sketch into Stable Diffusion 3.5 \citep{stability2024introducing-sd3.5}. Coupled with this model is IP-Adapter~\citep{ye2023ipadaptertextcompatibleimage} that helps generate the background in the style of the reference image, and ControlNet~\citep{zhang2023addingconditionalcontroltexttoimage} that generates the background from the user sketch. Given the size of the model, we generate a low pixel count image (512x384) for a fast controllable first generation pass. This image is then used to guide an AWS Bedrock-hosted model (Amazon Nova Canvas)~\citep{agi2025amazonnovafamilymodels} for fast high-quality, high resolution image variation generation, which is rescaled to desired dimensions. This was a faster and more robust approach than generating the high resolution image directly from the local large Stable Diffusion model. The Muse is then composited at a random location along the bottom axis of the final image.

\textbf{Generation task \#2: our muses in custom poses \& garments}

\begin{figure}[!h]
  \centering
  % \fbox{\rule[-.5cm]{0cm}{4cm} \rule[-.5cm]{4cm}{0cm}}
  \includegraphics[scale =0.4]{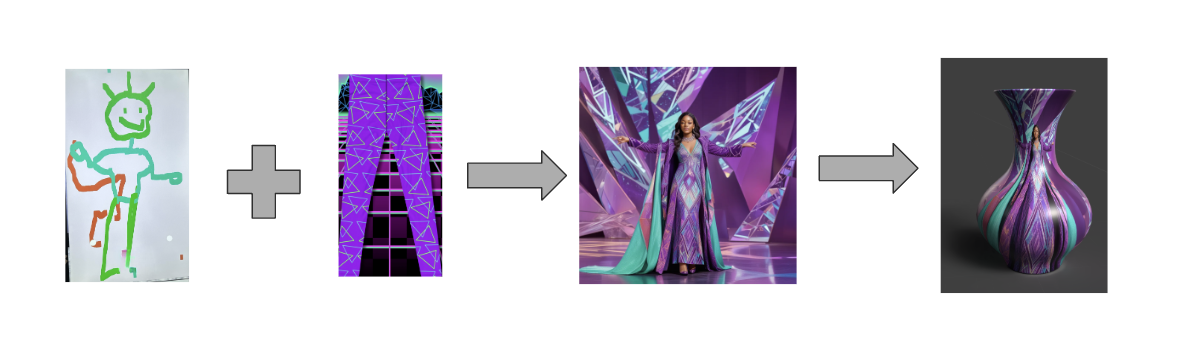}
  \caption{Generation Task 2, our muses in custom poses and garments}
\end{figure}

\textit{User Experience: } In this generation task, audience members were invited to sketch a pose for their assigned muse. Each muse, depicted in the chosen pose, was dressed with a designer specific garment style, which is then layered as a texture on various objects in Unreal Engine. The styles of designer garment (including colors, motifs, and patterns) were provided by the costume design team to ensure the garment each Muse was generated in was appropriate and fitting to the costumes the actors were seen in onstage. Acolytes guided audiences through this task, while the muses offered suggestions for poses they would like to be seen in.

\textit{Implementation: }We designed a multi-agent system to handle both audience and designer inputs, powered by Claude 3.5 Sonnet \citep{anthropic2024claude3.5-sonnet}. The first agent generates a detailed text description about the designer garment style. The second generates a text description of the pose from the audience sketch. The third one converts this text description into numerical pose keypoints. The latter two agents used few-shot learning in their prompting. The generated keypoints were processed to ensure body proportions were accurate to the muse. We then use the Instant ID framework~\citep{wang2024instantidzeroshotidentitypreservinggeneration}, which consists of the YamerMIX-8 model \citep{yamermix_Civitai}, a Stable Diffusion XL (SDXL) checkpoint~\citep{podell2023sdxlimprovinglatentdiffusion}, as the base diffusion model, Identity Net (IN) for facial preservation, and a Pose Control Net (PCN)~\citep{zhang2023addingconditionalcontroltexttoimage} to generate the image. The pose keypoints are passed into the PCN, the muse image is passed into IN, and the garment text description is passed directly into SDXL. In order to represent poses and faces as accurately as possible, we had the IN active throughout the entire generation, and the PCN start right after and end midway in the generation, capturing the essence of the desired pose while avoiding deleterious effects on facial features. For pose sketches that are too abstract, we created a custom pose library to randomly select a pose for the muse to appear in. From here, we generate an image of the muse in the specified garment and pose. This image is converted into a texture to be overlayed on various vases in the Unreal environment. 

\textbf{Generation task \#3:  3D object offerings for the muses}

\begin{figure}[!h]
  \centering
  % \fbox{\rule[-.5cm]{0cm}{4cm} \rule[-.5cm]{4cm}{0cm}}
  \includegraphics[scale =0.35]{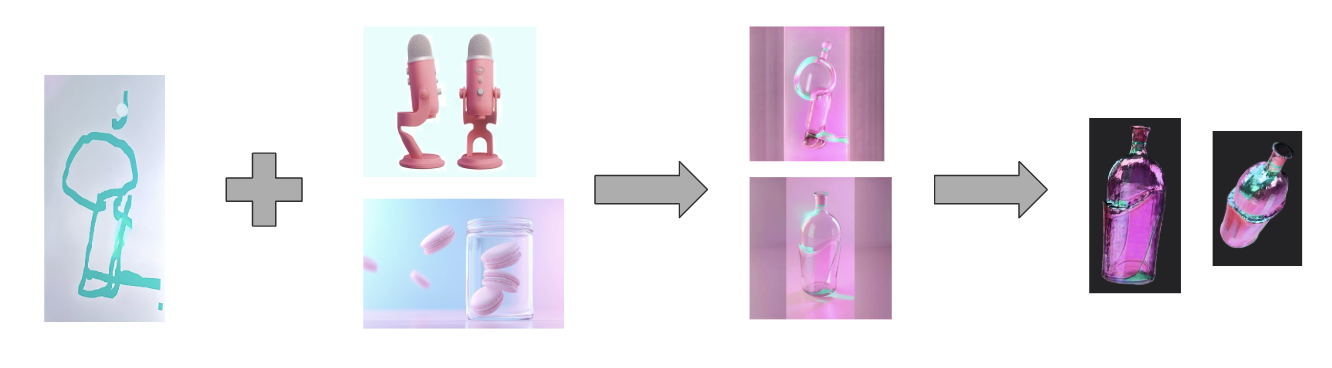}
  \caption{Generation Task 3, 3D object offerings for the muses}
\end{figure}

\textit{User Experience: }In this generation task, audience members drew sketches of various objects as offerings to the muses. The final output is a 3D asset of the object rendered back into the Unreal-powered world. The inputs for this generation are the audience sketch, along with a designer style reference. Designers provided vaporwave aesthetic reference images to ensure the final generations fit into \textit{Xanadu}'s 1980s retro vaporwave style. As objects rendered back onto the stage, actors improvised pointing out particularly interesting generations to audience members. 

\textit{Implementation: } The audience sketch is first passed into a VLM to generate a text description of the object. This description, along with the user sketch and a designer style reference, is passed to the SDXL diffusion model~\citep{podell2023sdxlimprovinglatentdiffusion} augmented with a content-style disentanglement module~\citep{xing2024csgocontentstylecompositiontexttoimage}, enabling the generation of a cleaner and more refined sketch that captures the original user sketch in the designer style. This then guides a larger diffusion model hosted on AWS Bedrock to rapidly generate a high quality image. Finally, the image is converted to a 3D mesh using an image-to-3D diffusion model~\citep{huang2025spar3dstablepointawarereconstruction} and rendered back on stage.

\subsection{System implementation}\label{appendix:compute}

In order to scale these generations to audiences of up to sixty-five every night, we developed a modular and scalable system architecture using Amazon Web Services. Drawings were first made on textures in the Unreal Engine virtual world through the show’s custom WebAR application, which sends data via Google Firebase to the game engine. Local Python code exported the drawings, and uploaded them to AWS Simple Storage Service (S3), an object storage service. This then notifies our serverless orchestration code hosted on AWS Lambda, a compute service that runs code in response to events without any external server provisioning.

Notifications are sent via Amazon Simple Notification Service (SNS) that publishes messages to Amazon Simple Queue Services (SQS), which, under the orchestration of a Lambda Helper Function, routed the incoming sketches to one of three generation modules used in the show. Each module generated media based on the drawings it received, invoking both hyperscale foundation models on AWS Bedrock, a service with access to multiple large language or image models, and smaller faster models deployed as real-time endpoints on Amazon SageMaker, AWS’s platform for building, training, and deploying models. The resulting generations were stored in S3 and fetched for human-in-the-loop moderation before being loaded into the Unreal Engine virtual world experienced by the audience. This process, from sketch to display, completes within 30-60 seconds, which fit within the show’s timing to ensure seamless audience participation.

Twenty-four SageMaker AI endpoints handled inference requests, eight for each generation task. We used 8 g6.12xlarge (48 vCPUs, 4x24GB Nvidia L4 GPUs) for generation tasks 1 and 3, and 16 g6.4xlarge (16 vCPUs, one 24GB Nvidia L4 GPU) instances for generation task 2. Each instance hosted a customized stable diffusion model and included support from Amazon Bedrock for average inference times of about 20-30 seconds for generation tasks 1 and 3, and about 40-60 seconds for generation task 2.

\subsection{Oracle: generated choreography}\label{appendix:oracle}

We developed a multimodal computer vision pipeline that integrates LLMs and pose estimation to generate and evaluate dance choreography. Initiated during live performances, the pipeline employed the GPT-4o model~\citep{openai2024gpt4ocard}, to process a randomly selected subset of audience-generated sketches collected each evening. A predefined list of twelve dance moves, derived from the musical number performed in that scene, provided the movement vocabulary for the interaction. Each night, the AI model selected three moves from this list and generated a short poem inspired by those selections, integrating the chosen gestures into its language to inform the Oracle's choreographic interpretation.

The second phase of the pipeline involved pose estimation and skeletal tracking using Stereolabs' ZED cameras. During playtesting, we used the ZED SDK's built-in 2D skeletal tracking system and evaluated its performance against ground-truth data captured from the Oracle's dance movements. The system tracked individual movement, allowing for comparative analysis across three key metrics: movement accuracy, timing, and dancing energy. Movement accuracy and timing were quantified using Object Keypoint Similarity (OKS) and Dynamic Time Warping (DTW), which were computed by analyzing 2D skeletal frames captured in real time and comparing them to the Oracle’s ground-truth choreography. Dancing energy was inferred from the velocity of joint movements over time, with greater joint velocity indicating higher levels of engagement and intensity. 

To make this data expressive and actionable within the performance, the computed metrics were normalized to a 0–1 scale and integrated into Unreal Engine. This mapping drove visual and environmental feedback, such as pulsing lights and visible progress in constructing the mythical \textit{Xanadu} auditorium, across both the virtual world and the physical set. In this way, this real-time feedback loop created a dramaturgical bridge between audience participation and narrative progression.

While the full pipeline demonstrated strong performance during playtesting with small audience groups, we encountered scalability limitations with the ZED SDK’s 2D skeletal tracking in larger crowd settings, where tracking fidelity declined. As a result, the pose estimation component was replaced with human-driven visual evaluation for production. Nevertheless, the pipeline provided valuable insights into the relationship between audience motion and theatrical response, offering a promising foundation for future work in responsive performance design and audience-aware dramaturgy.

\newpage
\section*{NeurIPS Paper Checklist}

\begin{enumerate}

\item {\bf Claims}
    \item[] Question: Do the main claims made in the abstract and introduction accurately reflect the paper's contributions and scope?
    \item[] Answer: \answerYes{} % Replace by \answerYes{}, \answerNo{}, or \answerNA{}.
    \item[] Justification: Yes, the abstract details the main contributions this paper makes, including a new ritual logic framework of audience input as offerings, a meta-instrument design pattern for a theater performance, and a reciprocal loop. Overall, we propose new contributions for group interfaces versus single user paradigms.
    % \item[] Guidelines:
    % \begin{itemize}
    %     \item The answer NA means that the abstract and introduction do not include the claims made in the paper.
    %     \item The abstract and/or introduction should clearly state the claims made, including the contributions made in the paper and important assumptions and limitations. A No or NA answer to this question will not be perceived well by the reviewers. 
    %     \item The claims made should match theoretical and experimental results, and reflect how much the results can be expected to generalize to other settings. 
    %     \item It is fine to include aspirational goals as motivation as long as it is clear that these goals are not attained by the paper. 
    % \end{itemize}

\item {\bf Limitations}
    \item[] Question: Does the paper discuss the limitations of the work performed by the authors?
    \item[] Answer: \answerYes{} % Replace by \answerYes{}, \answerNo{}, or \answerNA{}.
    \item[] Justification: Yes, please refer to \ref{subsec:limitations}. 

\item {\bf Theory assumptions and proofs}
    \item[] Question: For each theoretical result, does the paper provide the full set of assumptions and a complete (and correct) proof?
    \item[] Answer: \answerNA{} % Replace by \answerYes{}, \answerNo{}, or \answerNA{}.
    \item[] Justification: This paper does not include theoretical results of the types referenced in this item.
    % \item[] Guidelines:
    % \begin{itemize}
    %     \item The answer NA means that the paper does not include theoretical results. 
    %     \item All the theorems, formulas, and proofs in the paper should be numbered and cross-referenced.
    %     \item All assumptions should be clearly stated or referenced in the statement of any theorems.
    %     \item The proofs can either appear in the main paper or the supplemental material, but if they appear in the supplemental material, the authors are encouraged to provide a short proof sketch to provide intuition. 
    %     \item Inversely, any informal proof provided in the core of the paper should be complemented by formal proofs provided in appendix or supplemental material.
    %     \item Theorems and Lemmas that the proof relies upon should be properly referenced. 
    % \end{itemize}

    \item {\bf Experimental result reproducibility}
    \item[] Question: Does the paper fully disclose all the information needed to reproduce the main experimental results of the paper to the extent that it affects the main claims and/or conclusions of the paper (regardless of whether the code and data are provided or not)?
    \item[] Answer: \answerNA{} % Replace by \answerYes{}, \answerNo{}, or \answerNA{}.
    \item[] Justification: While this paper does describe an experiment--the production of \textit{Xanadu}--it is not designed as or intended to be reproducible.  Playtests, rehearsals, and proof-of-concepts were similarly bespoke. 

\item {\bf Open access to data and code}
    \item[] Question: Does the paper provide open access to the data and code, with sufficient instructions to faithfully reproduce the main experimental results, as described in supplemental material?
    \item[] Answer: \answerNA{} % Replace by \answerYes{}, \answerNo{}, or \answerNA{}.
    \item[] Justification: We have included the link to our Github repo in our appendix. This has majority of the code with additional  modules' code available upon request. However, as mentioned previously, \textit{Xanadu} is not intended to be reproducible, hence users may encounter varying results from the ones presented here.

\item {\bf Experimental setting/details}
    \item[] Question: Does the paper specify all the training and test details (e.g., data splits, hyperparameters, how they were chosen, type of optimizer, etc.) necessary to understand the results?
    \item[] Answer: \answerYes{} % Replace by \answerYes{}, \answerNo{}, or \answerNA{}.
    \item[] Justification: The experimental setting is presented in the paper with the description of various generative AI pipelines used for the sketching interaction in the show as well as the technology behind the computer vision pipelines powering the Oracle sequence. Please refer to \ref{appendix:sketching} and \ref{appendix:oracle} for further details on these workflows.
    %Naisha, Chiheb - should we include more technical information in a one to two page supplemental section if allowed? Not sure how much detail here
    % \item[] Guidelines:
    % \begin{itemize}
    %     \item The answer NA means that the paper does not include experiments.
    %     \item The experimental setting should be presented in the core of the paper to a level of detail that is necessary to appreciate the results and make sense of them.
    %     \item The full details can be provided either with the code, in appendix, or as supplemental material.
    % \end{itemize}

\item {\bf Experiment statistical significance}
    \item[] Question: Does the paper report error bars suitably and correctly defined or other appropriate information about the statistical significance of the experiments?
    \item[] Answer: \answerNA{} % Replace by \answerYes{}, \answerNo{}, or \answerNA{}.
    \item[] Justification:The paper does not include experimental results that can be analyzed statistically.
    % \item[] Guidelines:  The paper does not include experiment
    % \begin{itemize}
    %     \item The answer NA means that the paper does not include experiments.
    %     \item The authors should answer "Yes" if the results are accompanied by error bars, confidence intervals, or statistical significance tests, at least for the experiments that support the main claims of the paper.
    %     \item The factors of variability that the error bars are capturing should be clearly stated (for example, train/test split, initialization, random drawing of some parameter, or overall run with given experimental conditions).
    %     \item The method for calculating the error bars should be explained (closed form formula, call to a library function, bootstrap, etc.)
    %     \item The assumptions made should be given (e.g., Normally distributed errors).
    %     \item It should be clear whether the error bar is the standard deviation or the standard error of the mean.
    %     \item It is OK to report 1-sigma error bars, but one should state it. The authors should preferably report a 2-sigma error bar than state that they have a 96\% CI, if the hypothesis of Normality of errors is not verified.
    %     \item For asymmetric distributions, the authors should be careful not to show in tables or figures symmetric error bars that would yield results that are out of range (e.g. negative error rates).
    %     \item If error bars are reported in tables or plots, The authors should explain in the text how they were calculated and reference the corresponding figures or tables in the text.
    % \end{itemize}

\item {\bf Experiments compute resources}
    \item[] Question: For each experiment, does the paper provide sufficient information on the computer resources (type of compute workers, memory, time of execution) needed to reproduce the experiments?
    \item[] Answer: \answerYes{} % Replace by \answerYes{}, \answerNo{}, or \answerNA{}.
    \item[] Justification: Refer to \ref{appendix:compute} for information regarding compute.
    %Naisha, Chiheb - should we include more technical information in a one to two page supplemental section if allowed? Not sure how much detail here
    % \item[] Guidelines:
    % \begin{itemize}
    %     \item The answer NA means that the paper does not include experiments.
    %     \item The paper should indicate the type of compute workers CPU or GPU, internal cluster, or cloud provider, including relevant memory and storage.
    %     \item The paper should provide the amount of compute required for each of the individual experimental runs as well as estimate the total compute. 
    %     \item The paper should disclose whether the full research project required more compute than the experiments reported in the paper (e.g., preliminary or failed experiments that didn't make it into the paper). 
    % \end{itemize}
    
\item {\bf Code of ethics}
    \item[] Question: Does the research conducted in the paper conform, in every respect, with the NeurIPS Code of Ethics \url{https://neurips.cc/public/EthicsGuidelines}?
    \item[] Answer: \answerYes{} % Replace by \answerYes{}, \answerNo{}, or \answerNA{}.
    \item[] Justification: This research follows the NeurIPS Code of Ethics. Regarding potential harms in the research process, in particular with direct interactions with human participants, all audience members were informed when purchasing a ticket for \textit{Xanadu} about the tech elements at work in the show. The tech team worked with performers to ensure all audience members felt comfortable and not obliged to participate in the various interactions in the show if not interested/able. In regards to data, no dataset was directly used to train any models; all models used are released publicly and open for usage.

    % For societal impact and potential harms, the technology developed here is not harmful to audience members using it. Audience members were not recorded without their knowing and consent; for the show and dress rehearsal that was recorded, audience members had to sign a waiver acknowledging their consent. This can be harmful for the environment due to the GPUs being used from Amazon Web Services. In terms of bias and fairness, there are some patterns noticed in the second sketching module where actor's faces are being generated. We noticed some patterns with genAI generating certain races and genders that could potentially be harmful.
    
    % Naisha - review the code of ethics and explain some of how we informed the audience, worked with performers, etc. Jeff will look at IRB/HSR requirements and discussed below
    % \item[] Guidelines:
    % \begin{itemize}
    %     \item The answer NA means that the authors have not reviewed the NeurIPS Code of Ethics.
    %     \item If the authors answer No, they should explain the special circumstances that require a deviation from the Code of Ethics.
    %     \item The authors should make sure to preserve anonymity (e.g., if there is a special consideration due to laws or regulations in their jurisdiction).
    % \end{itemize}

\item {\bf Broader impacts}
    \item[] Question: Does the paper discuss both potential positive societal impacts and negative societal impacts of the work performed?
    \item[] Answer: \answerYes{} % Replace by \answerYes{}, \answerNo{}, or \answerNA{}.
    \item[] Justification: Please refer to \ref{subsec:limitations} for details on societal impact of our work (both positive and negative).

\item {\bf Safeguards}
    \item[] Question: Does the paper describe safeguards that have been put in place for responsible release of data or models that have a high risk for misuse (e.g., pretrained language models, image generators, or scraped datasets)?
    \item[] Answer: \answerNA{} % Replace by \answerYes{}, \answerNo{}, or \answerNA{}.
    \item[] Justification: All models used in this paper are pre-trained language models and diffusion models. During the show, operators in the crew managed generated content through a human in the loop system where images were checked before appearing on stage. In cases where images were deemed inappropriate, a fallback image was displayed instead. Please refer to \ref{subsec:invitation} for more details.
    %  Naisha, Chiheb - probably we have no such risks, but please consider and describe, updating paper / answer if needed]
    % \item[] Guidelines:
    % \begin{itemize}
    %     \item The answer NA means that the paper poses no such risks.
    %     \item Released models that have a high risk for misuse or dual-use should be released with necessary safeguards to allow for controlled use of the model, for example by requiring that users adhere to usage guidelines or restrictions to access the model or implementing safety filters. 
    %     \item Datasets that have been scraped from the Internet could pose safety risks. The authors should describe how they avoided releasing unsafe images.
    %     \item We recognize that providing effective safeguards is challenging, and many papers do not require this, but we encourage authors to take this into account and make a best faith effort.
    % \end{itemize}

\item {\bf Licenses for existing assets}
    \item[] Question: Are the creators or original owners of assets (e.g., code, data, models), used in the paper, properly credited and are the license and terms of use explicitly mentioned and properly respected?
    \item[] Answer: \answerYes{} % Replace by \answerYes{}, \answerNo{}, or \answerNA{}.
    \item[] Justification: Yes, all models and services used are cited (please refer to \ref{appendix} for details on each of these). There was no fine-tuning, and the use of other assets was consistent with typical approaches and license requirements for our productions.
    %[ Naisha, Chiheb - please update paper if needed with model / service cites (let's discuss design references for fine-tuning) and update here if needed. Not sure how much detail is needed her.]
    % \item[] Guidelines:
    % \begin{itemize}
    %     \item The answer NA means that the paper does not use existing assets.
    %     \item The authors should cite the original paper that produced the code package or dataset.
    %     \item The authors should state which version of the asset is used and, if possible, include a URL.
    %     \item The name of the license (e.g., CC-BY 4.0) should be included for each asset.
    %     \item For scraped data from a particular source (e.g., website), the copyright and terms of service of that source should be provided.
    %     \item If assets are released, the license, copyright information, and terms of use in the package should be provided. For popular datasets, \url{paperswithcode.com/datasets} has curated licenses for some datasets. Their licensing guide can help determine the license of a dataset.
    %     \item For existing datasets that are re-packaged, both the original license and the license of the derived asset (if it has changed) should be provided.
    %     \item If this information is not available online, the authors are encouraged to reach out to the asset's creators.
    % \end{itemize}

\item {\bf New assets}
    \item[] Question: Are new assets introduced in the paper well documented and is the documentation provided alongside the assets?
    \item[] Answer: \answerNA{} % Replace by \answerYes{}, \answerNo{}, or \answerNA{}.
    \item[] Justification: No new assets are released in this paper; \textit{Xanadu} is primarily an art piece that uses various AI components together in the context of a live performance.
    %[Naisha, Chiheb - I don't know if we have such assets? - something like: This is a short paper for the Creative AI track, primarily describing an artwork using various AI components in a large system. ]
    % \item[] Guidelines:
    % \begin{itemize}
    %     \item The answer NA means that the paper does not release new assets.
    %     \item Researchers should communicate the details of the dataset/code/model as part of their submissions via structured templates. This includes details about training, license, limitations, etc. 
    %     \item The paper should discuss whether and how consent was obtained from people whose asset is used.
    %     \item At submission time, remember to anonymize your assets (if applicable). You can either create an anonymized URL or include an anonymized zip file.
    % \end{itemize}

\item {\bf Crowdsourcing and research with human subjects}
    \item[] Question: For crowdsourcing experiments and research with human subjects, does the paper include the full text of instructions given to participants and screenshots, if applicable, as well as details about compensation (if any)? 
    \item[] Answer: \answerNA{} % Replace by \answerYes{}, \answerNo{}, or \answerNA{}.
    \item[] Justification: Please see also the note below regarding human subjects research. If appropriate for the Creative AI track, we can provide details on information provided to the audience ahead of the production for the camera ready version. 
    % \item[] Guidelines:
    % \begin{itemize}
    %     \item The answer NA means that the paper does not involve crowdsourcing nor research with human subjects.
    %     \item Including this information in the supplemental material is fine, but if the main contribution of the paper involves human subjects, then as much detail as possible should be included in the main paper. 
    %     \item According to the NeurIPS Code of Ethics, workers involved in data collection, curation, or other labor should be paid at least the minimum wage in the country of the data collector. 
    % \end{itemize}

\item {\bf Institutional review board (IRB) approvals or equivalent for research with human subjects}
    \item[] Question: Does the paper describe potential risks incurred by study participants, whether such risks were disclosed to the subjects, and whether Institutional Review Board (IRB) approvals (or an equivalent approval/review based on the requirements of your country or institution) were obtained?
    \item[] Answer: \answerNA{} % Replace by \answerYes{}, \answerNo{}, or \answerNA{}.
    \item[] Justification: Our theatrical productions are not typically subject to IRB review and we do not use data collected in the show's systems to draw generalizable conclusions about human behavior.  Discussion about audiences in this paper are based on our observations of public performances in which the audience has been informed of technology use and could opt out of each type of participation without consequence. 
    % \item[] Guidelines:
    % \begin{itemize}
    %     \item The answer NA means that the paper does not involve crowdsourcing nor research with human subjects.
    %     \item Depending on the country in which research is conducted, IRB approval (or equivalent) may be required for any human subjects research. If you obtained IRB approval, you should clearly state this in the paper. 
    %     \item We recognize that the procedures for this may vary significantly between institutions and locations, and we expect authors to adhere to the NeurIPS Code of Ethics and the guidelines for their institution. 
    %     \item For initial submissions, do not include any information that would break anonymity (if applicable), such as the institution conducting the review.
    % \end{itemize}

\item {\bf Declaration of LLM usage}
    \item[] Question: Does the paper describe the usage of LLMs if it is an important, original, or non-standard component of the core methods in this research? Note that if the LLM is used only for writing, editing, or formatting purposes and does not impact the core methodology, scientific rigorousness, or originality of the research, declaration is not required.
    %this research? 
    \item[] Answer: \answerYes{} % Replace by \answerYes{}, \answerNo{}, or \answerNA{}.
    \item[] Justification: Yes; the usage of LLMs is described as a part of the experimental procedures in this paper and as a part of the core methodology. Please refer to \ref{appendix} for more details on which LLMs were used and how they were used in the various pipelines in the show.
    % \item[] Guidelines:
    % \begin{itemize}
    %     \item The answer NA means that the core method development in this research does not involve LLMs as any important, original, or non-standard components.
    %     \item Please refer to our LLM policy (\url{https://neurips.cc/Conferences/2025/LLM}) for what should or should not be described.
    % \end{itemize}

\end{enumerate}

\end{document}